\documentclass[aps,prb,twocolumn,groupedaddress]{revtex4-1}

\usepackage{graphicx}
\usepackage{amsmath}
\usepackage{amsfonts}
\usepackage{amssymb}
\usepackage{array}
\usepackage{color}

\let\vec\mathbf
\newcommand{\hatb}[1]{\boldsymbol{\hat{\mathbf{#1}}}}

\begin{document}

\title{Intermediate phase and pseudo phase transition in an artificial spin ice model}

\author{R.A. Stancioli}
\author{L.A.S. M\'{o}l}
\email{lucasmol@fisica.ufmg.br}
\affiliation{Laborat\'orio de Simula\c {c}\~ao, Departamento de F\'isica, ICEx, Universidade Federal de Minas Gerais, 31270-901, Belo Horizonte, Minas Gerais, Brazil}

\date{\today}

\begin{abstract}
In this paper we conduct Monte Carlo simulations to investigate the thermodynamic properties of a geometry of artificial spin ice recently proposed in the literature that had been termed ``rewritable'' spin ice, for its experimental realization allows total control over the microstates of the system. Our results show that in the thermodynamic limit a single phase transition between a fully magnetized state and a paramagnetic state exists, whereas for finite systems an intermediate phase also emerges, engendering a low temperature pseudo phase transition. This intermediate phase is characterized by large magnetic domains separated by domain walls composed of monopole-like excitations, resulting in low net magnetization values. We also show that two types of low energy excitations that behave as magnetic monopoles emerge in the system, both of which are geometrically constrained to move along a predefined path.
\end{abstract}

\maketitle

\section{Introduction}

Artificial spin ices (ASI) are arrays of interacting magnetic nanoislands that can be litographically patterned in a variety of lattice structures, the most common examples being the square \cite{Wang2006} and the kagome \cite{Tanaka2006} lattices. The islands are small enough that they are single-domain, and the elongated shape of each individual island constrains its magnetic moment to point along its longest axis, forcing it to behave as an effective Ising macrospin. In the simplest ASI realization, the square ASI, the lowest energy configurations of each vertex of the lattice, where four spins meet, are characterized by two spins pointing inward and the other two directed outward, obeying an ice rule analogous to the one encountered in some water ice phases \cite{Bernal1933,Pauling1935}.

Initially designed to allow the real-space observation of the exotic states found in geometrically frustrated rare-earth pyrochlore crystals \cite{Harris1997, Bramwell2001}, ASI's scope has significantly transcended their original purpose, mainly due to their remarkable property of allowing the design of custom lattice structures \cite{Nisoli2013,Nisoli2017}
which deliberately target the emergence of behaviors and applications. In addition to the intriguing monopole-like excitations \cite{Mol2009a} which are also encountered in their three-dimensional (3D) counterparts \cite{Castelnovo2008}, ASI have opened windows to the study of novel phenomena such as vertex frustration \cite{Morrison2013}, monopole-charge screening \cite{Gilbert2014,Farhan2016}, topologically protected magnetic charges
\cite{Morrison2013,Lao2018}, transfer and processing of information \cite{Arava2018}, among others.

One of the most promising applications of these materials is in the storage and processing of information through the manipulation of their local configurations. In this regard, an important step has been made by Wang \textit{et al.} \cite{Wang2016}, who have designed a lattice geometry derived from the original square artificial spin ice that allows complete experimental control over the microstates of the system. For this reason, the system has been termed rewritable artificial spin ice (RWASI). Interesting is the fact that in this geometry the same monopole-like excitations picture of the original ASI is expected. However, due to the lattice structure, the monopoles are restricted to move along straight lines, suggesting the possibility of dimensionality reduction\cite{Gilbert2015}.

The experimental realization of this particular geometry was essentially athermal, since the relatively large volume of the nanoislands used in Ref. \onlinecite{Wang2016} posed a significant energy barrier for the inversion of spin orientations, blocking the system's dynamics. However, the fabrication of thinner nanoislands with a much lower blocking temperature is also possible \cite{Kapaklis2012}, and has been applied to a number of different spin ice geometries \cite{Porro2013,Farhan2013,Gilbert2015,Shi2017,Gliga2017,Arnalds2016}, shedding light on the thermodynamics of this class of materials both in and out of equilibrium\cite{Farhan2013a,Farhan2019}.

In order to elucidate the thermal behavior of this geometry of ASI, we have conducted Monte Carlo simulations of the RWASI using open (OBC) and periodic boundary conditions (PBC). Our results indicate that in the thermodynamic limit the system is expected to behave as the square ASI\cite{Nascimento2012}, exhibiting a phase transition in the Ising universality class between the ordered low temperature phase and the disordered high temperature phase. Nevertheless, for finite systems a pseudo phase transition appears at low temperature, marking the passing from the fully ordered (magnetized) phase to a low magnetization phase characterized by the presence of large domains separated by domain walls composed of magnetic monopole-like excitations. 
The appearance of this intermediate phase in finite systems makes for interesting possibilities, since the pseudotransition between the magnetized and low magnetization states is very abrupt, resembling a discontinuous phase transition, in such a way that small temperature variations may completely change the system's state and properties. The paper is organized as follows: Sec. II describes our model and simulation methods; in Sec. III, an analysis of vertices types and monopole excitations is carried out; Sec. IV is devoted to the system with periodic boundary conditions; Sec. V contains the results for open boundary conditions, and in Sec. VI we present our final remarks.

\section{Model and simulations}

\begin{figure}[!ht]
	\centering
    \includegraphics[width=.4\linewidth]{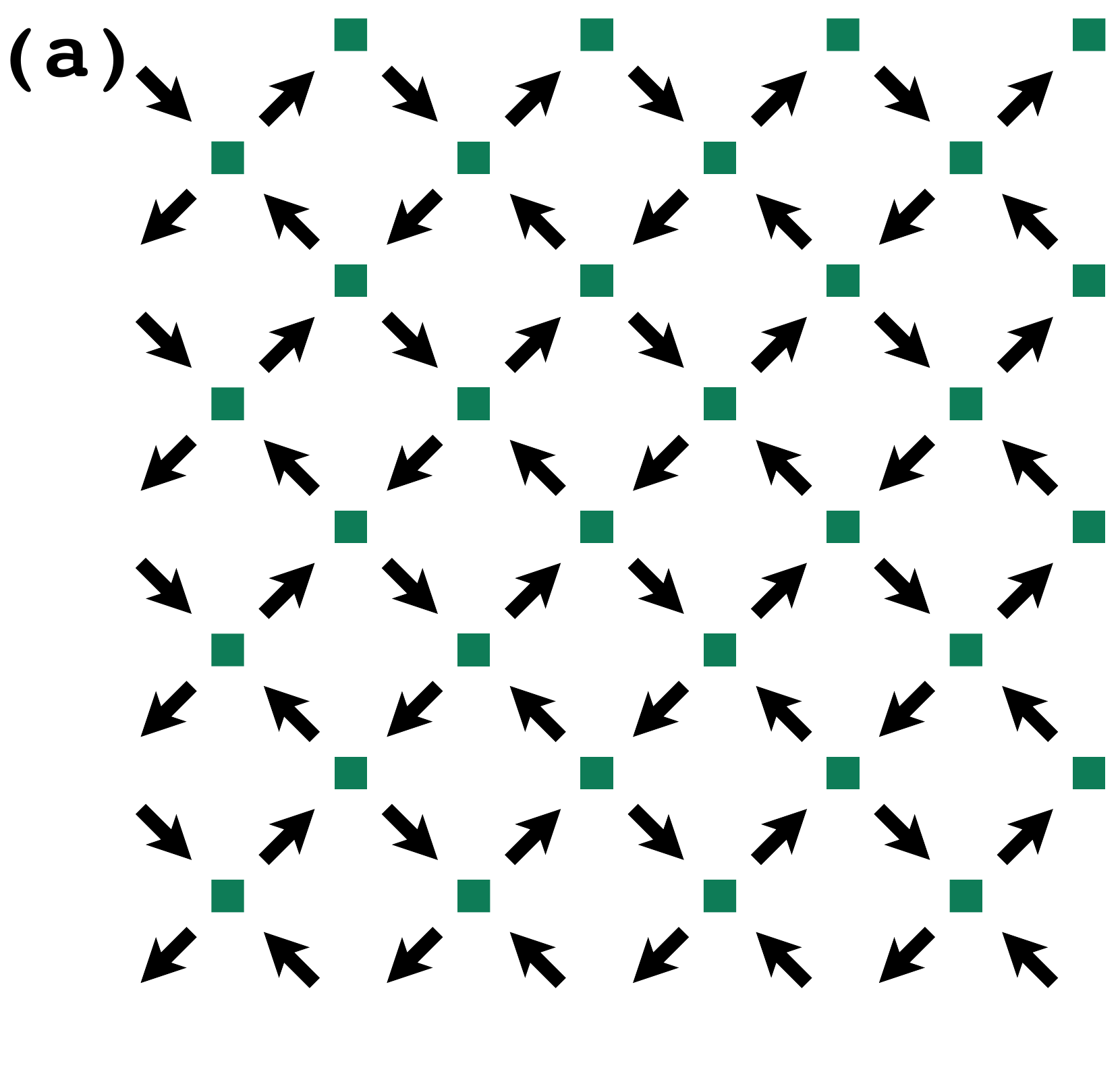}
\hspace{.1\linewidth}    \includegraphics[width=.4\linewidth]{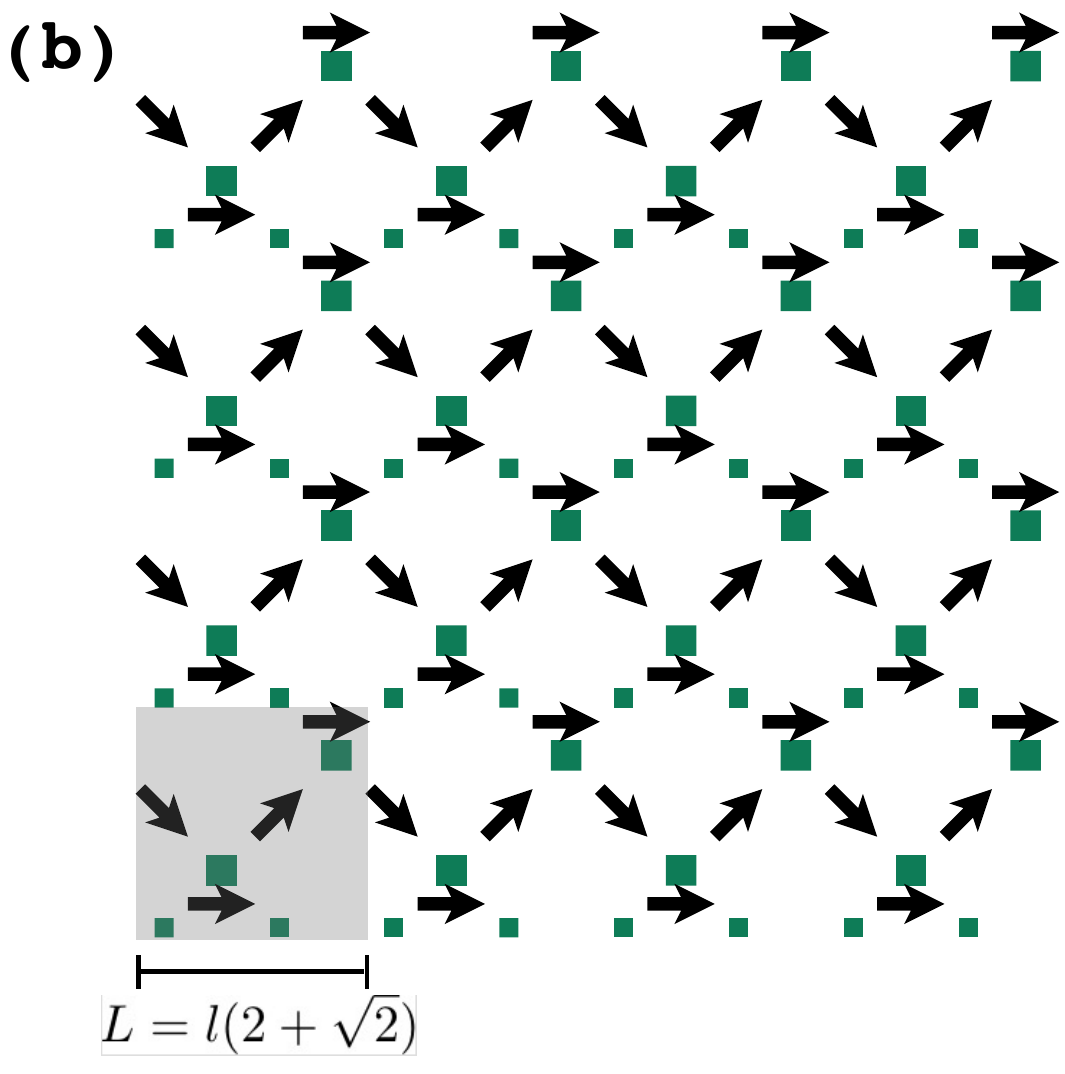}
    \caption{\textbf{(a)} A typical square artificial spin ice. \textbf{(b)} The RWASI geometry is obtained by modifying half of the spins in the square lattice. To do so, the dipoles of the square ASI are thought of as being composed of two poles, each of which is reattached to the opposite pole of a neighboring spin, resulting in rows of horizontal spins at a 45 deg angle with the original ones. The vertices of the square ASI (green squares) are still present in the modified lattice, but in the latter only two spins converge toward each of these vertices. Notice also that the rows of horizontal spins in the RWASI geometry give rise to secondary vertices between pairs of slightly misaligned dipoles.}
    \label{fig:lattice}
\end{figure}

The RWASI lattice is obtained by modifying the square ASI geometry as indicated in Fig. \ref{fig:lattice}. In the latter, each vertex is surrounded by two pairs of spins aligned in perpendicular directions, whereas in the former two of those spins are now placed at a 45 deg angle with respect to the others \cite{Wang2016}. As a result, each vertex of the RWASI lattice is surrounded by three spins, two of which are original, in the sense that they are also present in the square lattice. The third spin, which is the modified one, is placed at a 45 deg angle with respect to the other two. Half of the spins in the lattice are original and half are modified, since each original spin is shared between two vertices whereas the modified one is fully contained within a single plaquette. A more detailed discussion on vertex types will be offered in Sec. III.

In our model, nanoislands are treated as point-like Ising spins interacting through a dipolar potential. The magnetic moment of each island is given by $\vec{m}_i=\mu S_i \hatb{e}_i$, where $\mu$ is the norm of the dipole moment (considered equal for all nanoislands), $S_i=\pm1$ are the spin variables, and $\hatb{e}_i$ represents the fixed orientation of each spin on the plane. The lattice consists of $L \times L$ unit cells of four spins each, as shown in Fig. \ref{fig:lattice}(b), and the total number of spins is $N=4L^{2}$. The lattice parameter $l$ is defined in accordance with Ref. \onlinecite{Wang2016}, so that the separation between two consecutive unit cells is $l(2+\sqrt{2})$ in both $x$ and $y$ directions. The total Hamiltonian is written as:
\begin{equation}
\label{eq:hamil2}
\mathcal{H}=\frac{D}{2} \sum_{i \neq j} \left[ \frac{S_i \hatb{e}_i \cdot S_j \hatb{e}_j}{r_{ij}^3}-\frac{3 \left( S_i \hatb{e}_i \cdot \vec{r}_{ij} \right) \left( S_j \hatb{e}_j \cdot \vec{r}_{ij} \right)}{r_{ij}^5} \right],
\end{equation}
where $\vec{r}_{ij}$ is the vector connecting spins at sites $i$ and $j$, and $D=\mu_0 \mu^2/(4 \pi l^3)$ is the coupling constant of the dipolar interaction, $\mu_0$ being the permeability of free space.

We ran simulations with two different boundary conditions: open and periodic. In the lattice with open boundary conditions, the last spin in each row has been removed to avoid an asymmetry between left and right borders. For periodic boundary conditions, several copies of the system have been considered in each direction. As the dipolar potential in two dimensions is generally convergent, we assume that the amount of copies ($\sim10^5$) used in the simulations is large enough that the total energy can be computed with reasonable precision. This assumption was verified by comparing the results for simulations performed considering different cutoff radii
\footnote{Tests have been carried out for $L=8$ and $L=32$ with 100, 200 and 500 copies in each direction.}
and noticing that the results agree within error bars.

Since the only variable in each term of the summation in Eq. \ref{eq:hamil2} is the product $S_iS_j$, the Hamiltonian can be factored as
\begin{equation}
\label{eq:hamil3}
\mathcal{H}= \frac{D}{2} \sum_{i \neq j} S_i S_j \left[ \frac{\hatb{e}_i \cdot \hatb{e}_j}{r_{ij}^3}-\frac{3 \left(\hatb{e}_i \cdot \vec{r}_{ij} \right) \left(\hatb{e}_j \cdot \vec{r}_{ij} \right)}{r_{ij}^5} \right].
\end{equation}
By calculating the constant term in brackets in Eq. \ref{eq:hamil3} for each pair of spins beforehand, the computing time needed to update energy values during simulations is greatly reduced.
We used a single spin-flip Metropolis algorithm, combined with some multiple spin-flip steps to speed up the dynamics and avoid trapping in local minima. The results of the simulations were extrapolated to non-simulated temperatures through the multiple histogram reweighting method \cite{Ferrenberg1989}. For each lattice size and boundary condition, ten samples were simulated to allow the estimation of errorbars. A typical simulation consisted of $10^5$ Monte Carlo steps for equilibration and $10^6$ steps to compute averages.

\section{Vertices and monopoles}

\begin{figure*}[t]
	\centering
    \includegraphics[width=1.0\textwidth]{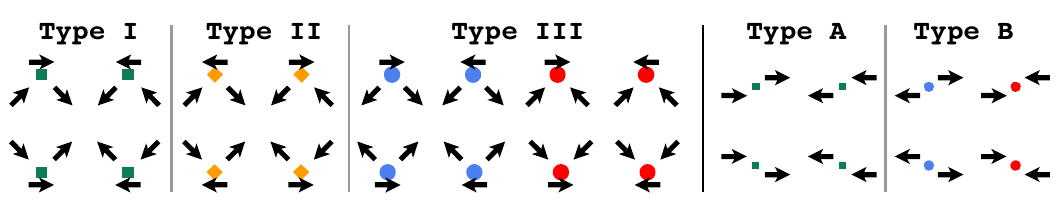}
    \caption{The eight possible configurations of the primary vertex are classified in three types, in ascending order of interaction energy. The vertices shown in the second row are in the same configuration as the vertices immediately above them, but they are reflected about the $x$ axis. In a random configuration of the system, the expected values of vertex populations are $25\%$ for types I and II and $50\%$ for type III vertices. The secondary vertices are not present in square ASI and are classified in types A and B. The head-to-tail alignment of spins in type A vertices makes them less energetic than type B.}
    \label{fig:vertices}
\end{figure*}

As can be noticed in Fig. \ref{fig:lattice}, each vertex of the square ASI has its counterpart in RWASI. However, the ice rule does not hold in the latter, at least in its original formulation. In the modified lattice, the net magnetic charge of a vertex is kept neutral whenever one of the original spins points in and the other one points out. The violation of this ice rule analog yields an excess of magnetic charge in the vertex, in a similar way to what happens in the square lattice when the ice rule is violated. The orientation of the modified spin is important for energy considerations but does not contribute to the magnetic charge in the vertex, because the two opposite poles of this spin are equidistant to the vertex.

Since a vertex is now defined by only three spins, it can assume eight possible configurations, which can be divided in three different groups according to their energy levels. It is important to notice, though, that these configurations may appear reflected about the horizontal axis, as shown in Fig. \ref{fig:vertices}, because the modified spin appears above the other two in half of the vertices and below them in the other half. Both type I and type II vertices obey the modified ice rule and are charge neutral, but type I vertices are less energetic than type II because all of its local interactions are satisfied. Type III vertices are the most energetic and exhibit an excess of magnetic charge. The doubly charged vertices present in square ASI do not occur in the modified lattice.

Contrary to what happens in square ASI, in the RWASI lattice not all spins point towards their respective vertices. Indeed, the modified spins in RWASI form structures very similar to one-dimensional Ising arrays, with a small offset in the $y$ direction between nearest neighbors. The midpoint between two of these spins can be treated a secondary vertex of coordination number $z=2$. Energy is minimized when both spins have the same orientation in the $x$ direction (type A vertices), whereas excited vertices (type B) appear when spins have opposite orientations (see Fig. \ref{fig:vertices}).

\begin{figure}[!h]
	\centering
    \includegraphics[width=0.45\linewidth]{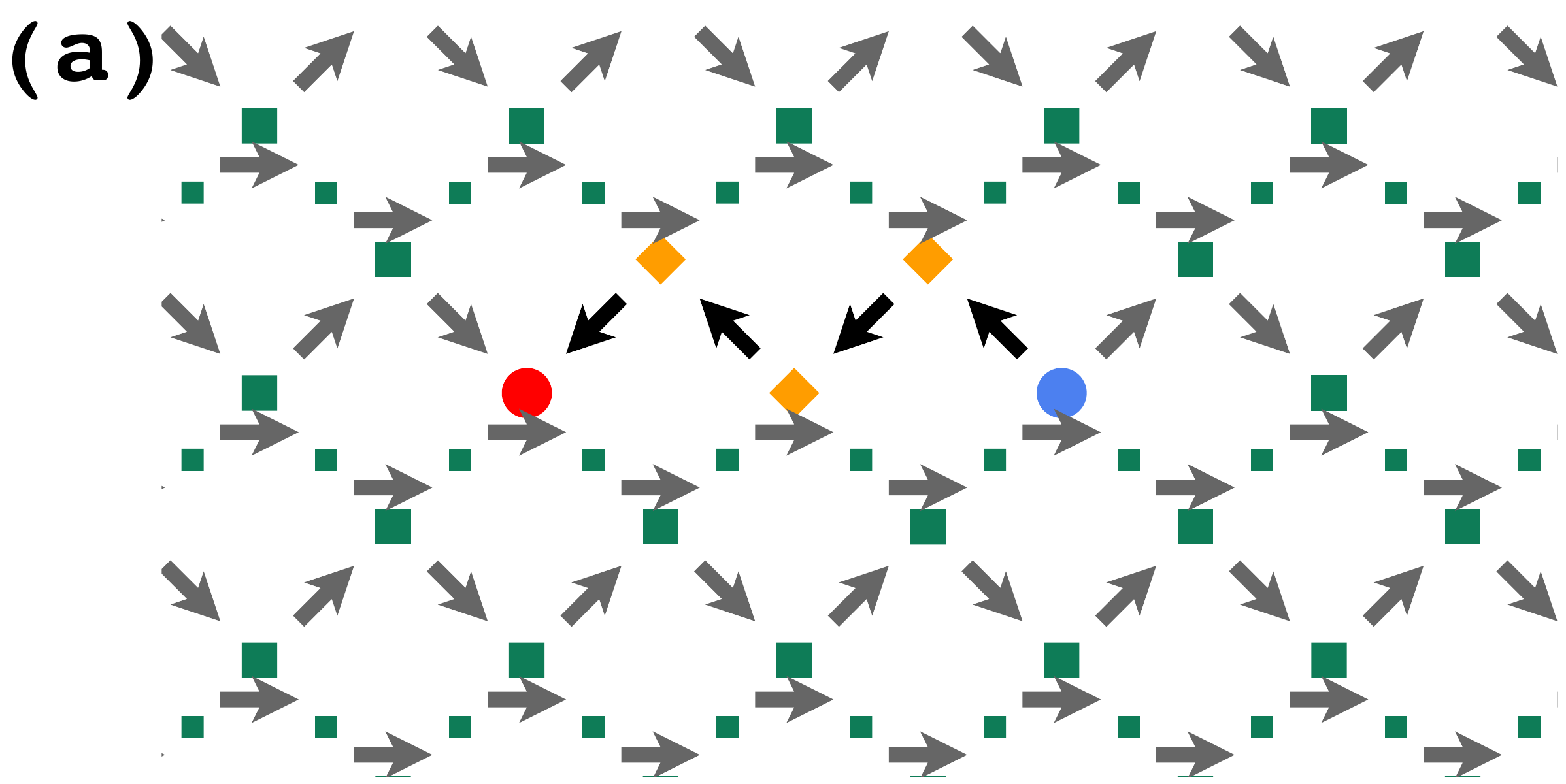}
    \includegraphics[width=0.45\linewidth]{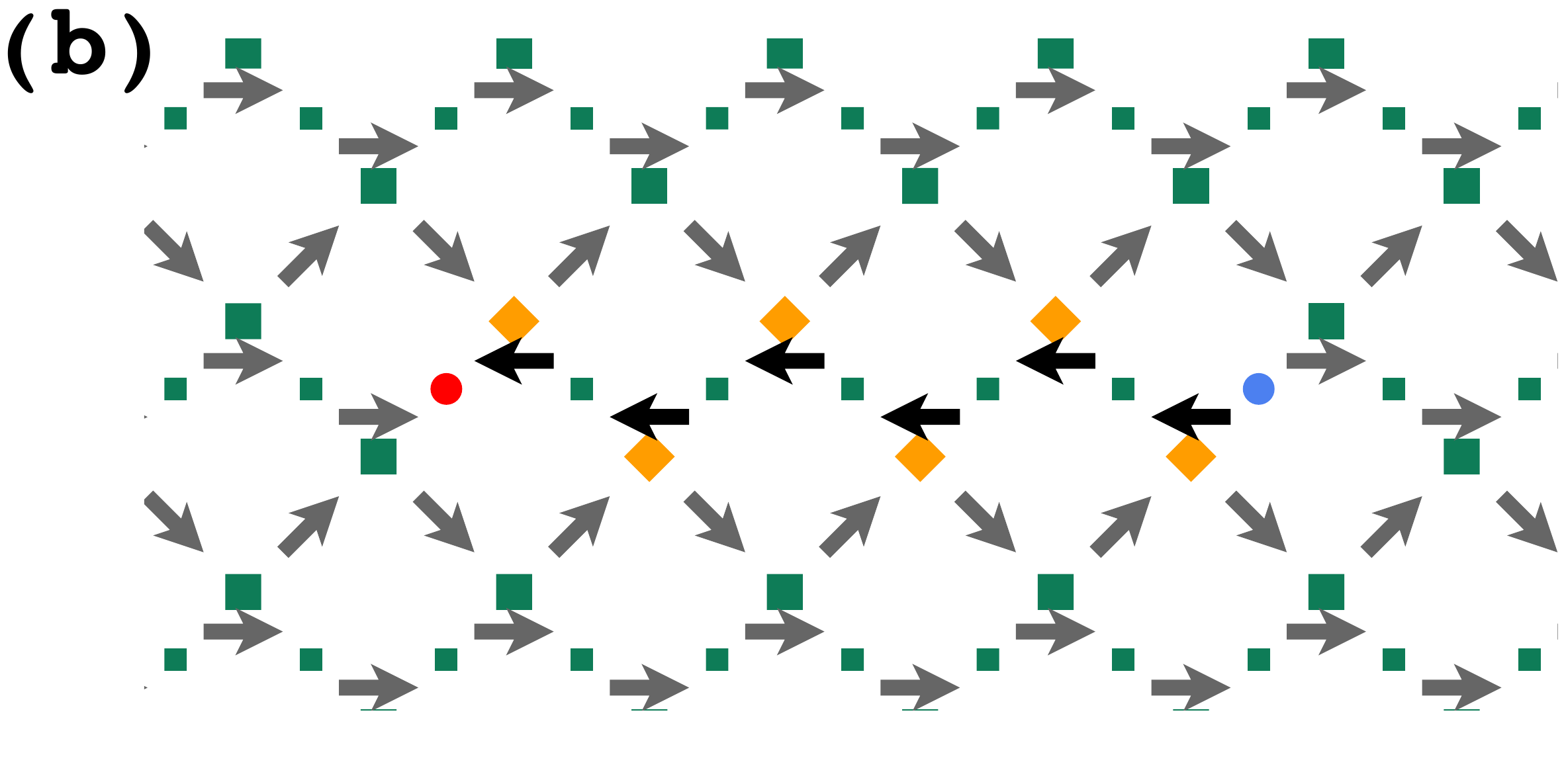}
    \caption{\textbf{(a)} A pair of oppositely charged type III excitations (circles) formed by flipping a sequence of spins (highlighted in black) from the ground state. Notice that by flipping neighboring spins one can only separate them in a zigzag path. \textbf{(b)} Two type B monopole-like excitations (smaller circles) formed by flipping a sequence of modified (horizontal) spins. These monopoles are constrained to move along a straight line. In both cases, the opposite poles are connected by a string of type II vertices (diamond-shaped markers).}
    \label{fig:monopoles}
\end{figure}

\begin{figure}[!h]
	\centering
    \includegraphics[width=\linewidth]{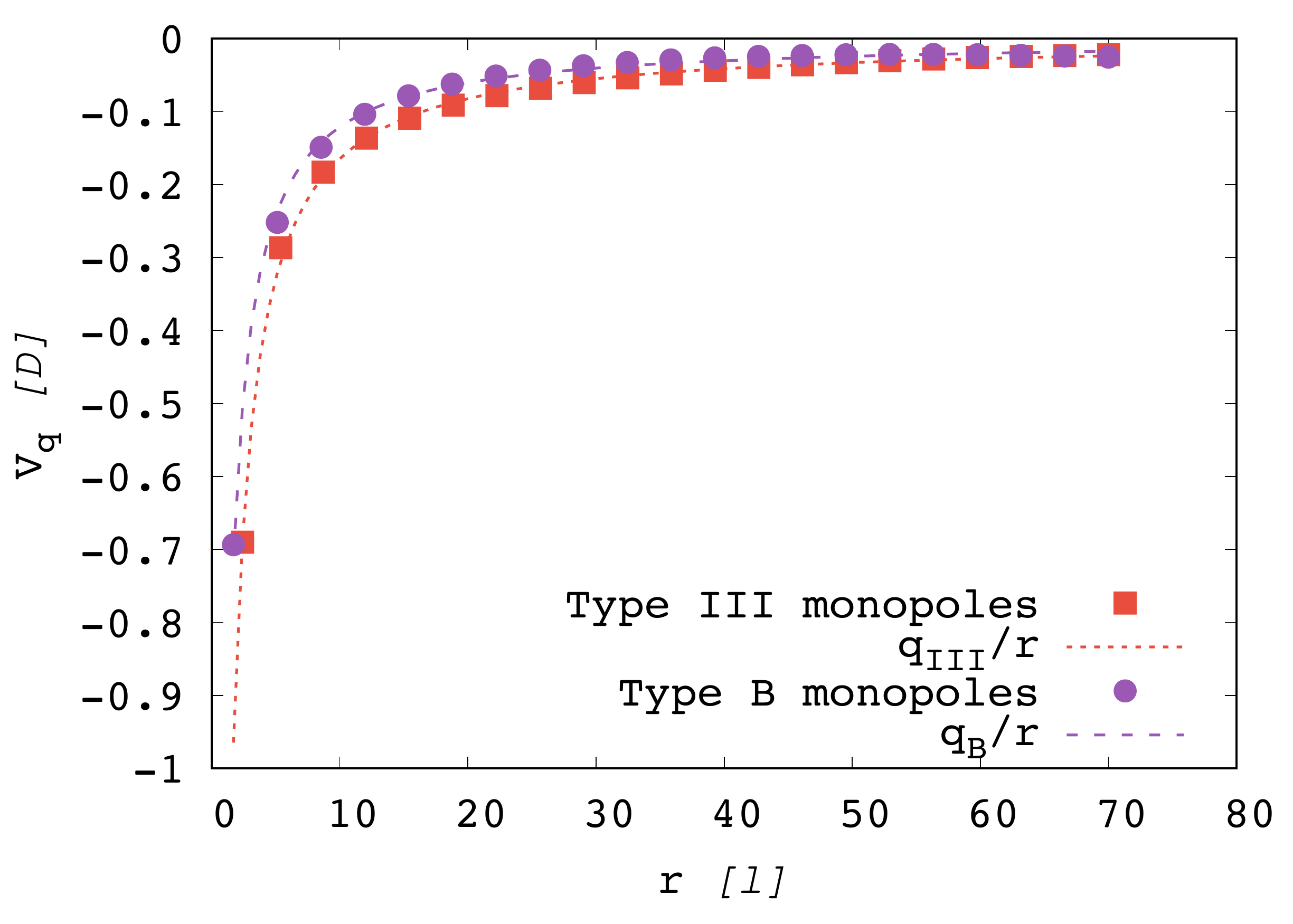}
    \caption{The Coulombic term $V_q \equiv q/r$ of Eq. \ref{eq:monopole} is plotted for the two types of monopole-like excitations present in the system. The magnetic charge of monopoles has been estimated by fitting Eq. \ref{eq:monopole} to the potential energy of monopoles separated by a distance $r$ (dashed lines). The charge $q_{III}$ of monopoles formed by the emergence of type III vertices was determined to be greater than the charge $q_{B}$, which correspond to type B vertices.}
    \label{fig:monopolos}
\end{figure}

In square ASI, it has been shown \cite{Mol2009a} that elementary excitations above the ground state appear as a pair of oppositely charged vertices, whose interaction energy exhibits an effective Coulombic term. The separation of the excited vertices across the lattice result in a string of charge-neutral yet energetically disfavored vertices. Thus, the energy of these excitations can be written as:
\begin{equation}
\label{eq:monopole}
V(r)= -\frac{q}{r} + ax(r) + b,
\end{equation}
where $q$ is the magnetic charge magnitude, $r$ is the distance between the excited vertices, $a$ is the string tension, $x(r)$ is the string length, and $b$ is the energy required for monopole pair formation. In the RWASI lattice, this behavior is observed not only for type III vertices, which are analogous to the single-charged monopoles of the square lattice, but also for type B vertices, that are specific to this geometry. In both cases, the separation of a pair of monopoles across the lattice results in a string of type II vertices, that are charge-neutral but considerably more energetic than type I (see Fig. \ref{fig:monopoles}). By adjusting Eq. \ref{eq:monopole} to the energy of excited configurations, we were able to calculate the effective magnetic charge of type III and type B vertices, which are, respectively, $q_{III}=1.65(2)Dl$ and $q_B=1.21(1)Dl$ (see Fig. \ref{fig:monopolos}).

One important consequence of this geometry is the fact that monopoles are now constrained to move along a predefined path. From Fig. \ref{fig:monopoles}(a), it is clear that a pair of type III monopoles created in the $i^{th}$ row of the lattice can only follow a zigzag path that either leads them further apart or brings them closer together in the same row until they meet and annihilate. The same happens with type B monopoles, which follow a straight line when hopping between secondary vertices (Fig. \ref{fig:monopoles}(b)). This allows for the possibility of controlled magnetic currents which is absent in conventional ASI arrays.

\section{Periodic boundary conditions}

\begin{figure}[!ht]
	\centering
		\includegraphics[width=0.45\textwidth]{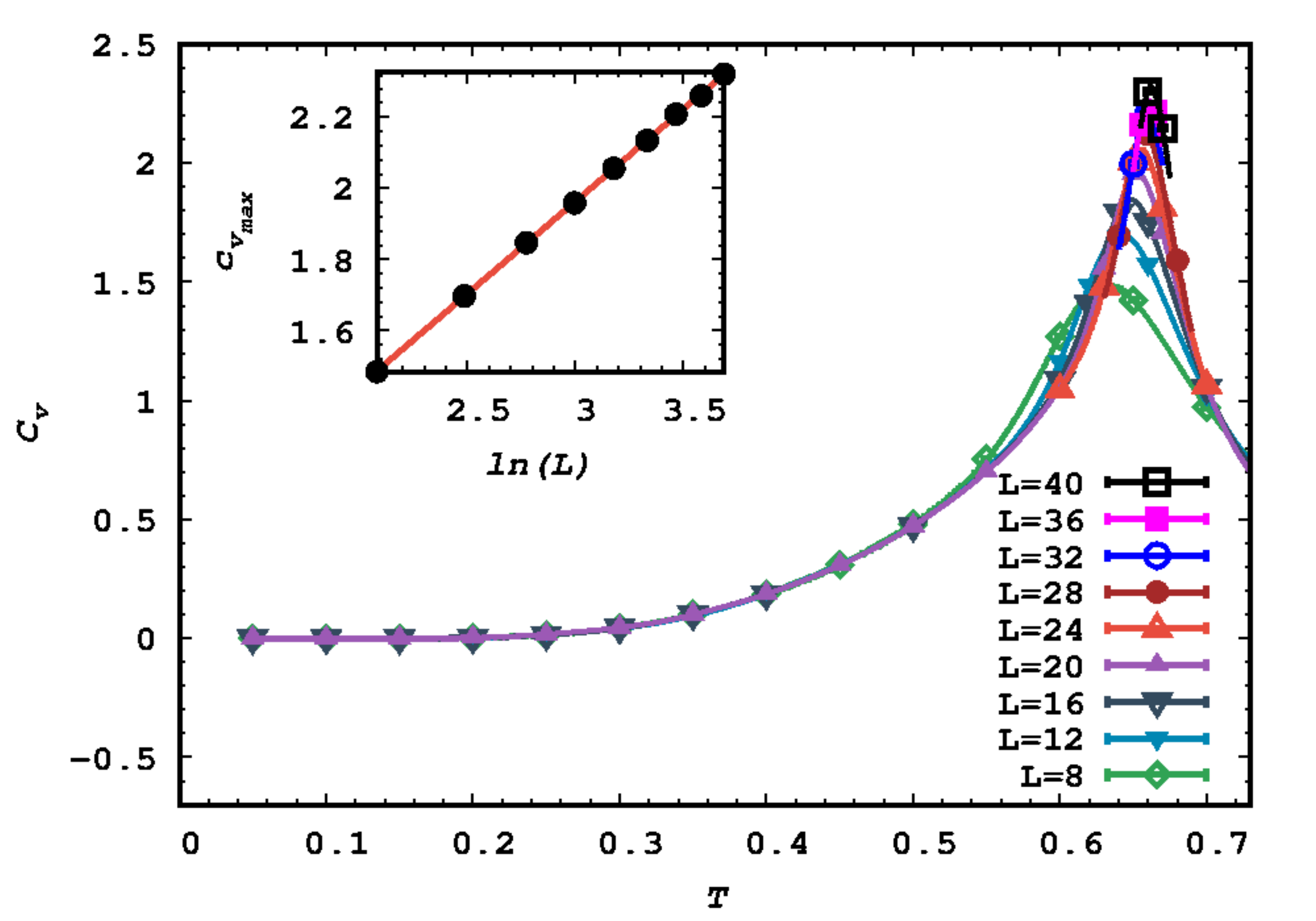}%
    \caption{ The specific heat curve for PBC exhibits a peak near the transition temperature that becomes more pronounced with increasing lattice sizes. \textbf{(inset)} The linear fit of the specific heat peak as a function of the natural logarithm of the lattice size $L$ suggests a logarithmic divergence, resulting in a critical exponent $\alpha =0$. The points in the graphs are the direct results of simulations, whereas the lines were obtained by the multiple histogram reweighting technique. When not shown error bars ares smaller than the symbol sizes.}
    \label{fig:pico_pbc}
\end{figure}

Contrary to the square ASI, whose ground state is characterized by an alternate tiling of type I vertices that results in no net magnetization \cite{Nisoli2007}, the ground state of the RWASI was determined to be fully magnetized. This is due to the fact that the RWASI geometry removes the frustration of the square lattice at the local level, since it allows a head-to-tail arrangement of all nearest-neighbor spins, as can be inferred from the configuration of its least energetic vertices. In the ground state, all vertices are in type I and type A configurations, as shown in Fig. \ref{fig:lattice}(b). As the temperature rises, the magnetization drops rapidly and a peak in the specific heat curve is observed (see Fig. \ref{fig:pico_pbc}) as the system transitions into a paramagnetic phase.

We simulated lattices of sizes ranging from $L=8$ to $L=40$ to investigate the nature of this phase transition. The finite-size scaling of the specific heat peak revealed a logarithmic divergence, as shown in the inset of Fig. \ref{fig:pico_pbc}, indicating that the critical exponent that governs the specific heat behavior, denoted by $\alpha$, equals zero for this transition. A power-law fit was also tried, but resulted in a worse fit (not shown). This is in accordance with previous results for the square ASI \cite{Silva2012}, where a logarithmic divergence was also found.

In order to estimate the critical temperature we applied a method based on the energy probability distribution (EPD) zeros recently reported in the literature\cite{Costa2017}. This method consists of a modification of the Fisher zeros \cite{Fisher1965} approach and allows the calculation of the complex zeros of the partition function from a single histogram constructed at a certain inverse temperature $\beta_0$. As in the Fisher zeros method, phase transitions are characterized by the zeros that touch the real positive axis in the thermodynamic limit. For finite systems, however, there can be no real positive zeros, so that one expects that the zeros nearest to the positive real axis, called leading or dominant zeros, are indicative of phase transitions. This method presents some advantages over conventional ones, such as not requiring the definition of an order parameter and allowing the obtention of the critical temperature directly from the partition function---and not from the behavior of derived thermodynamic quantities such as specific heat and susceptibility. In addition, an estimate of the transition exponent $\nu$ can also be easily obtained. More details about this method can be found in the Appendix and in Refs. \onlinecite{Costa2017,Mol2018,Costa2019}.

\begin{figure}[!ht]
		\centering
		\includegraphics[width=.45\textwidth]{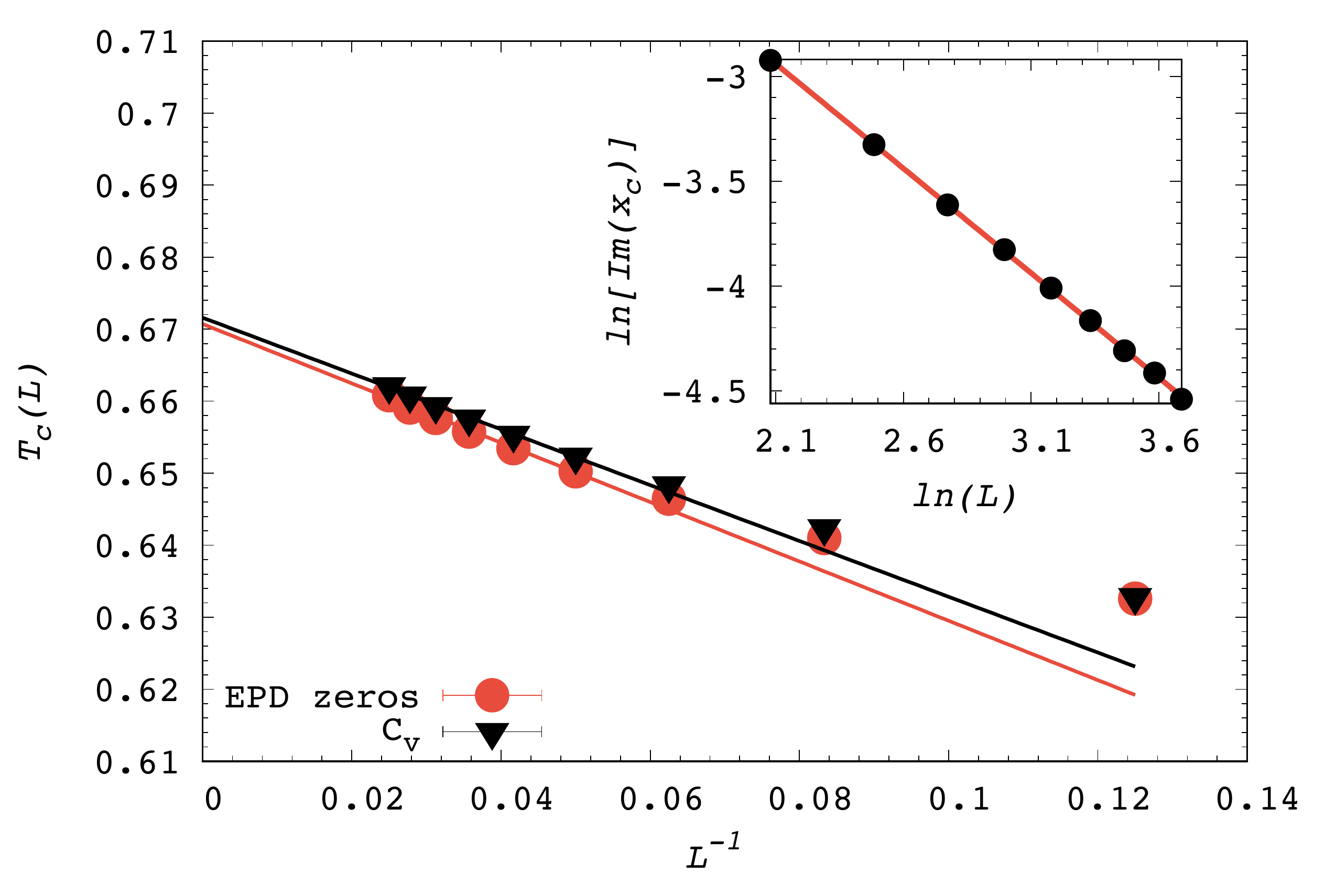}
	    \caption{Critical temperature obtained with both the EPD zeros method and the specific heat peak. The Ising exponent $\nu=1$ was used. The $y$ intercept of the fit gives the critical temperature in the thermodynamic limit. Smaller lattice sizes ($L=8, 12, 16$) were not considered in the fit. The inset shows a log-log plot of the imaginary part of the leading zero as a function of lattice size. The slope of the linear fit allows the estimation of the critical exponent $\nu=1.001(5)$. When not shown error bars ares smaller than the symbol sizes.}
	    \label{fig:temp_pbc}
\end{figure}

The inset of Fig. \ref{fig:temp_pbc} shows a log-log plot of the imaginary part of the leading zero as a function of lattice size. From Eq. \ref{eq:imag}, the slope of a linear fit on the data is equivalent to ${-1/\nu}$, which in this case yields the critical exponent $\nu=1.001(5)$. This result, along with the logarithmic divergence of the specific heat, indicates that this transition belongs to the two-dimensional Ising universality class, a fact that has been previously observed in models of square ASI \cite{Xie2015}.

We then used the critical temperature values obtained with finite lattices, $T_c(L)$, to estimate the critical temperature in the thermodynamic limit $T_c^*$, according to the expression
\begin{equation}
\label{eq:8.58}
T_c(L) = T_c^* + a L^{-1 / \nu}
\end{equation}
where $a$ is a constant. Taking the Ising exponent $\nu=1$ and plotting $T_c(L)$ as a function of $L^{-1}$ results in a straight line whose intercept with the $y$ axis gives the critical temperature $T_c^*$. Using the values of $T_c(L)$ given by the EPD zeros method, we found a critical temperature of $T_c^*=0.671(1)D/k_B$, while extracting $T_c(L)$ from the position of the specific heat peak yielded a very close value, $T_c^*=0.672(1)D/k_B$ (see Fig. \ref{fig:temp_pbc}). The critical temperature $T_c^*=0.671D/k_B$, along with the Ising exponents $\gamma=1.75$ and $\nu=1$, generates a good collapse of both magnetic susceptibility and Binder cumulant curves, providing a good consistency check for our results (not shown here).

The critical temperature of the square ASI has been determined by Silva \textit{et al.} \cite{Silva2012} to be approximately $T_c \approx 7.2D'/k_B$, where $D'=\mu_0 \mu^2/(4 \pi l'^3)$ is the coupling constant of the dipolar interaction used in that work and $l'$ is the lattice parameter. A comparison between the critical temperatures of the square ASI and the rewritable ASI is only possible if we notice that the definition of the lattice parameter $l$ used in our study is different from the lattice parameter $l'$ defined in Ref. \onlinecite{Silva2012}. While $l'$ represents the distance between two adjacent vertices in the square lattice, the corresponding vertices in the rewritable lattice are separated by a distance of $l(1+\sqrt{2})$.
Therefore, in order for the position of the corresponding vertices in both lattices to coincide, we should have $l=l'/(1+\sqrt{2})$, which yields a ratio between coupling constants of $D=(1+\sqrt{2})^3D'$. The critical temperature $T_c=0.671(1)D/k_B$ of the RWASI model would thus result in $T_c=9.44(1)D'/k_B$ in the units used in Ref. \onlinecite{Silva2012}, which is higher than that of the square ASI. A possible explanation is the higher level of frustration present in the square ASI, which makes ordering more difficult.

\section{Open boundary conditions}

\begin{figure}[!ht]
	\centering
    \includegraphics[width=.45\textwidth]{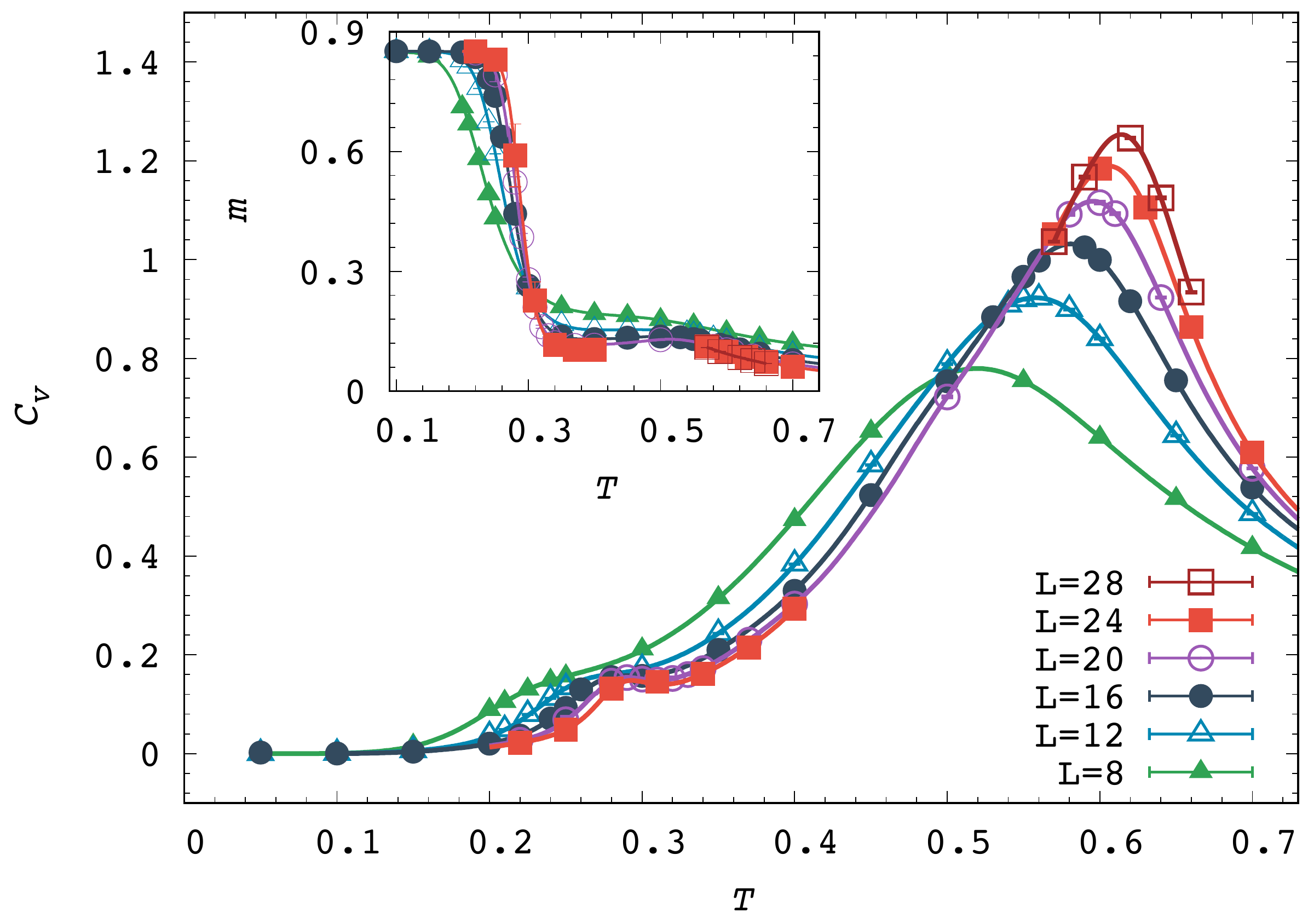}
	 \caption{The specific heat curve for OBC signals two significant regime changes, as a local maximum that is absent in PBC now appears. The magnetization curve (inset) shows that the temperature at which the system is fully magnetized is much lower for OBC than for PBC. Also, an intermediate regime in magnetization arises, and is identified by a plateau in the curve. The paramagnetic phase is reached at a slightly lower temperature when compared to PBC.}
    \label{fig:mc}
\end{figure}

\begin{figure}[!ht]
	\centering
    \includegraphics[width=.3\textwidth]{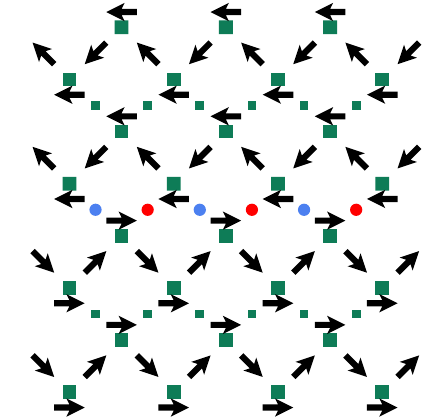}
	 \caption{The typical configuration of the intermediate phase exhibits a domain wall between two clusters of opposite magnetization, resulting in low values of total magnetization. This domain wall requires the emergence of type B and/or type II vertices. Here, a domain wall of type B vertices is shown, with positive and negative charge excitations represented as red (dark gray) and blue (light gray) circles respectively.}
    \label{fig:fase_inter}
\end{figure}

Remarkably, the system's behavior is different when open boundary conditions are used, suggesting that finite-size effects give rise to some features that become suppressed in the thermodynamic limit. From Fig. \ref{fig:mc} (inset), it is clear that the total magnetization exhibits three different regimes, and not only two as in the PBC case. The ground state is still maximally magnetized, as one would expect, but as the temperature rises the magnetization curve reaches a plateau before finally decaying into the paramagnetic phase. The specific heat curve is also qualitatively different, as another local maximum arises at a much lower temperature than the formerly observed peak (Fig. \ref{fig:mc}). This local maximum is significantly less pronounced for smaller lattices, and appears as a mere shoulder in the curve for $L=8$ and $L=12$. By looking at typical configurations for these three regimes, we notice that between the fully magnetized and the paramagnetic phases there is an intermediate phase that is characterized by the emergence of a domain wall of type B and type II vertices separating clusters of opposite spin orientations. These clusters, however, contain almost no magnetically charged vertices (Fig. \ref{fig:fase_inter}).

\begin{figure}[!ht]
	\centering
    \includegraphics[width=.45\textwidth]{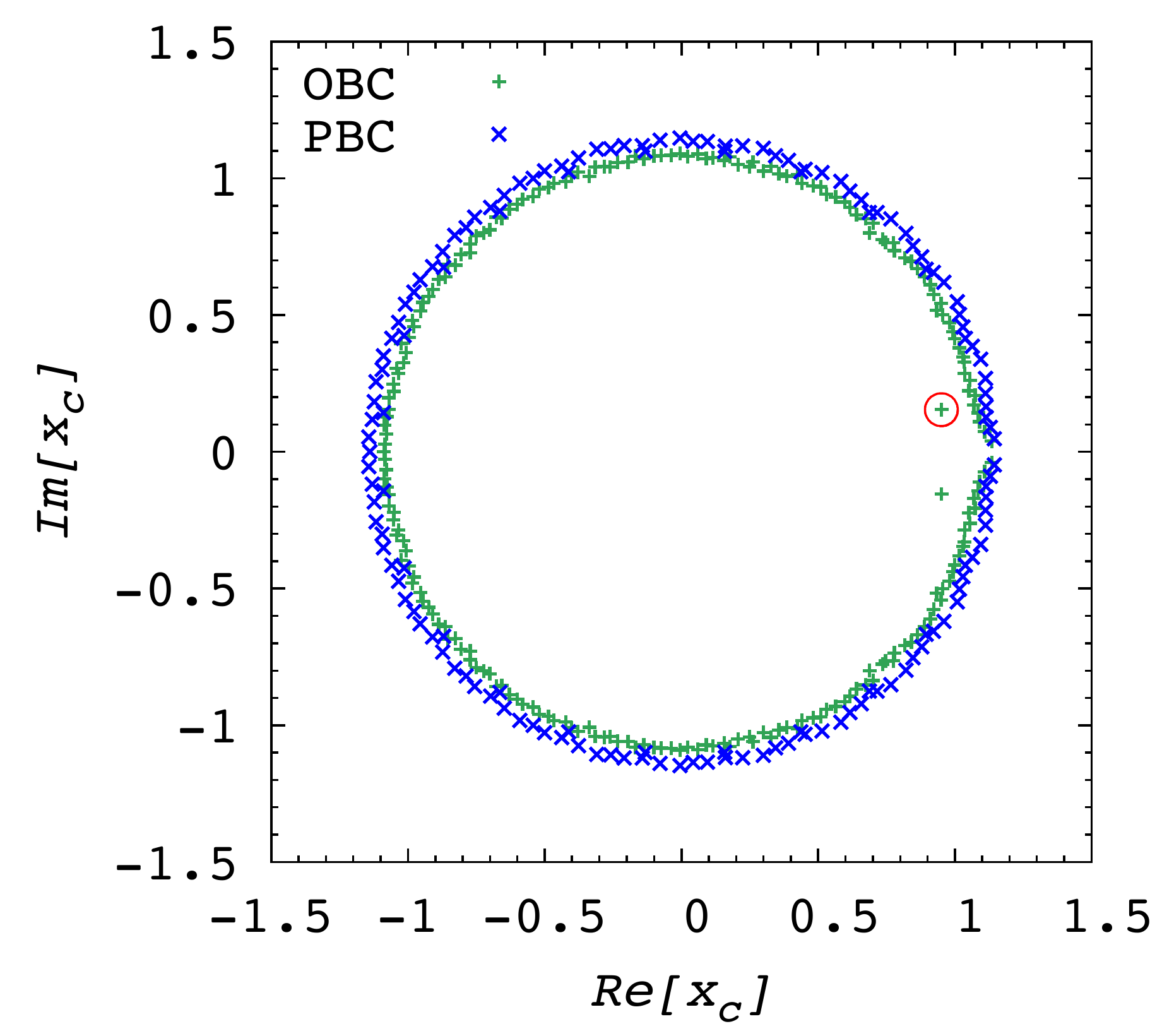}
    \caption{Comparison between the zeros maps for PBC and OBC lattices, both with $L=8$ and $T_0=1/\beta_0=0.27D/k_B$. The finite system presents two zeros that converge to a real part of one upon iteration. The zero highlighted in the map, that is clearly absent for the periodic lattice, leads to the low-temperature pseudo phase transition.}
    \label{fig:root_minor}
\end{figure}

Although phase transitions can only be rigorously defined in the thermodynamic limit, this behavior of the finite system suggests that it undergoes two pseudo phase transitions. 
This is further confirmed by the EPD zeros method. Fig. \ref{fig:root_minor} shows a comparison between the zeros maps for PBC and OBC lattices of the same size and at the same temperature. In the OBC lattice's map, two zeros clearly stand out, both of which converge to a real part of one upon iteration, indicating two pseudo phase transitions. The PBC lattice, on the other hand, presents only one dominant zero as expected.
We simulated OBC lattices ranging from $L=8$ to $L=28$. The temperatures at which these transitions occur, shown in Table \ref{tab:tc}, satisfactorily agree with the temperatures of the specific heat peaks for each system size.

\begin{table}[ht]
	\begin{center}
    \caption{\centering Temperature of pseudo phase transitions for different system sizes.}
    \label{tab:tc}
		\begin{tabular}{>{\centering\arraybackslash}m{1in}  >{\centering\arraybackslash}m{1in} >{\centering\arraybackslash}m{1in}}
				\cline{1-3}
				$L$& $T_{c_1} (D/k_B)$ & $T_{c_2} (D/k_B)$
				\\ \hline \hline
				$8$ &     $0.2439(1)$&	$0.569(1)$
				\\ 
				$12$ & 	$0.2645(1)$&	$0.588(1)$
				\\ 
				$16$ & 	$0.2765(7)$&	$0.600(2)$
				\\ 
				$20$ &	$0.2843(5)$&	$0.611(2)$
				\\ 
				$24$ &	$0.2884(4)$&	$0.620(3)$
				\\ 
				$28$ &	$0.293(2)$&	$0.622(2)$
				\\ \hline
		\end{tabular}%
	\end{center}
\end{table}

\begin{figure}[!ht]
	\centering
	\includegraphics[width=.9\linewidth]{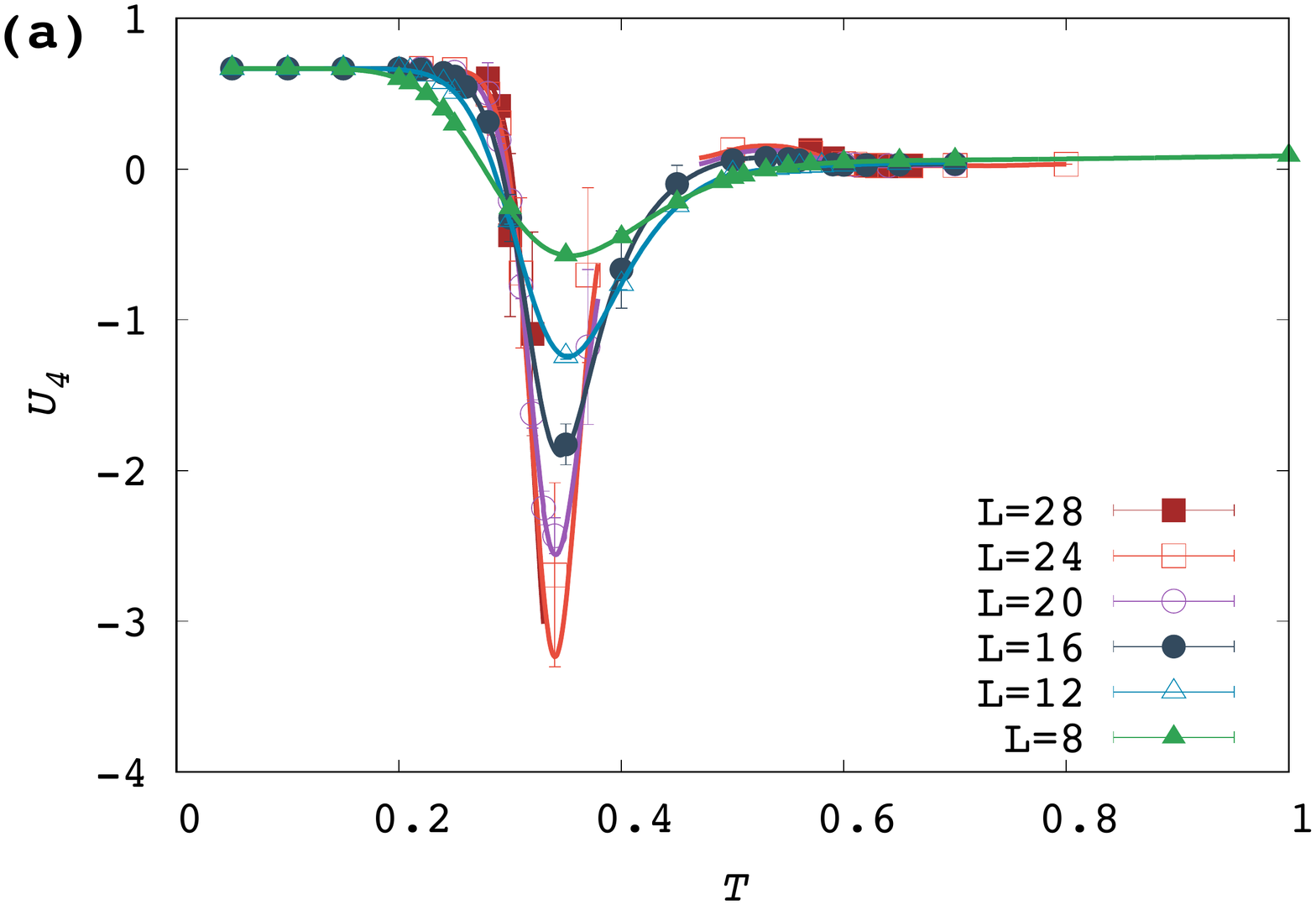}
	\includegraphics[width=.9\linewidth]{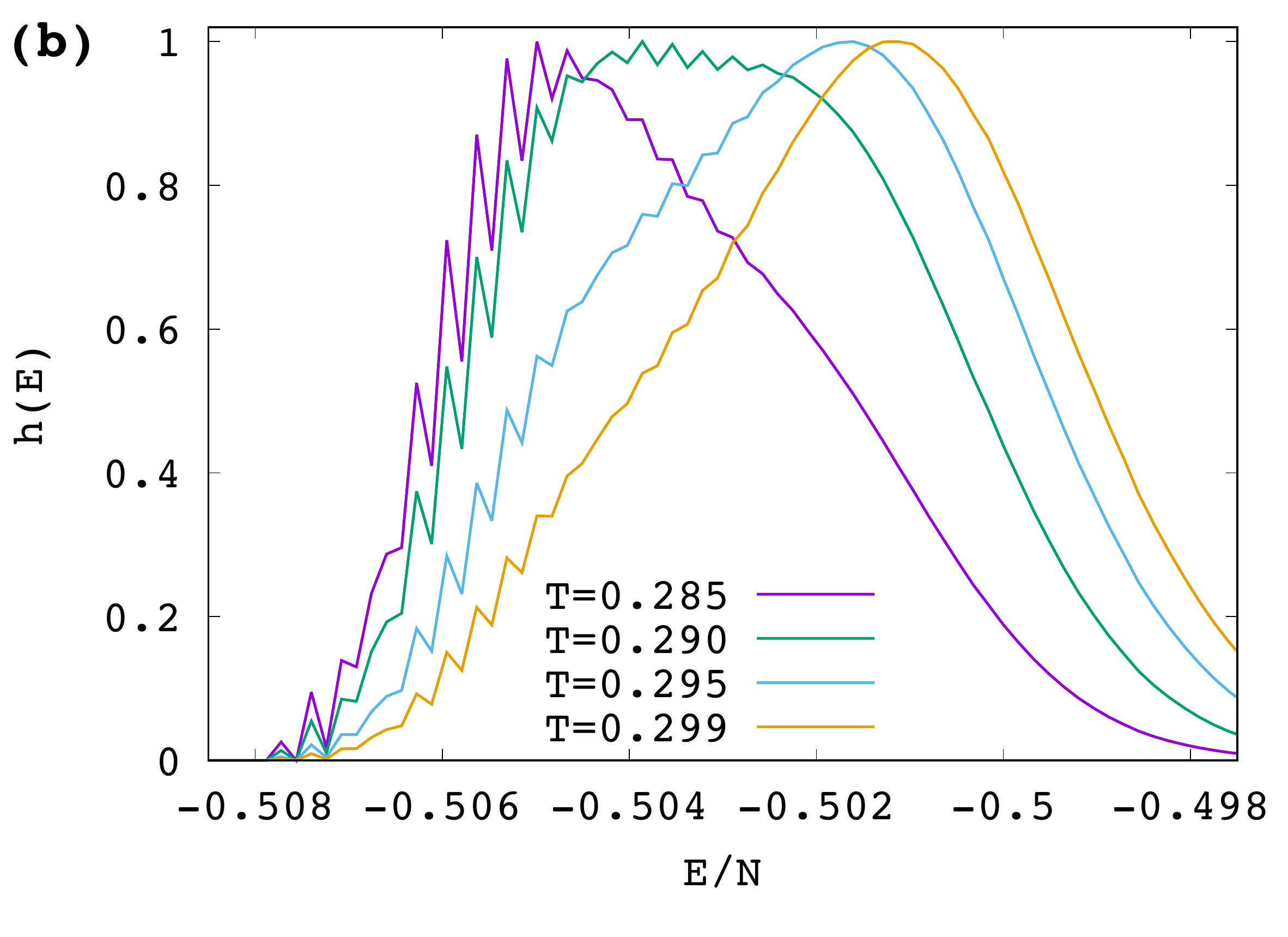}
    \caption{\textbf{(a)} Binder cumulant of the magnetization with OBC for different system sizes. The curves exhibit a negative minimum that sharpens with increasing lattice size. \textbf{(b)} Energy histograms with $L=28$, OBC. The pseudo phase transition for this lattice size occurs at $T_c=0.293(2)$. The rough appearance of the curve at low energies is due to the unevenly spaced energy levels of possible configurations, which are placed in equally-sized energy bins for the construction of the histograms.}
    \label{fig:binderobc}
\end{figure}

This low-temperature pseudo phase transition appears to be first order in nature. Indeed, the Binder cumulant of the magnetization,
$$U_4=1-\frac{<m^4>}{3<m^2>^2},$$ exhibits a sharp drop that begins near the transition temperature and reaches negative values (Fig. \ref{fig:binderobc}(a)), suggesting a first order transition \cite{Tsai1998}. Although the energy histogram near the transition is not clearly double-peaked, we can 
infer that it is formed by the sum of two unimodal distributions as can be seen in the histograms above and below the transition temperature (see Fig. \ref{fig:binderobc}(b)).

\begin{figure*}[t]
	\centering
	\includegraphics[width=.4\linewidth]{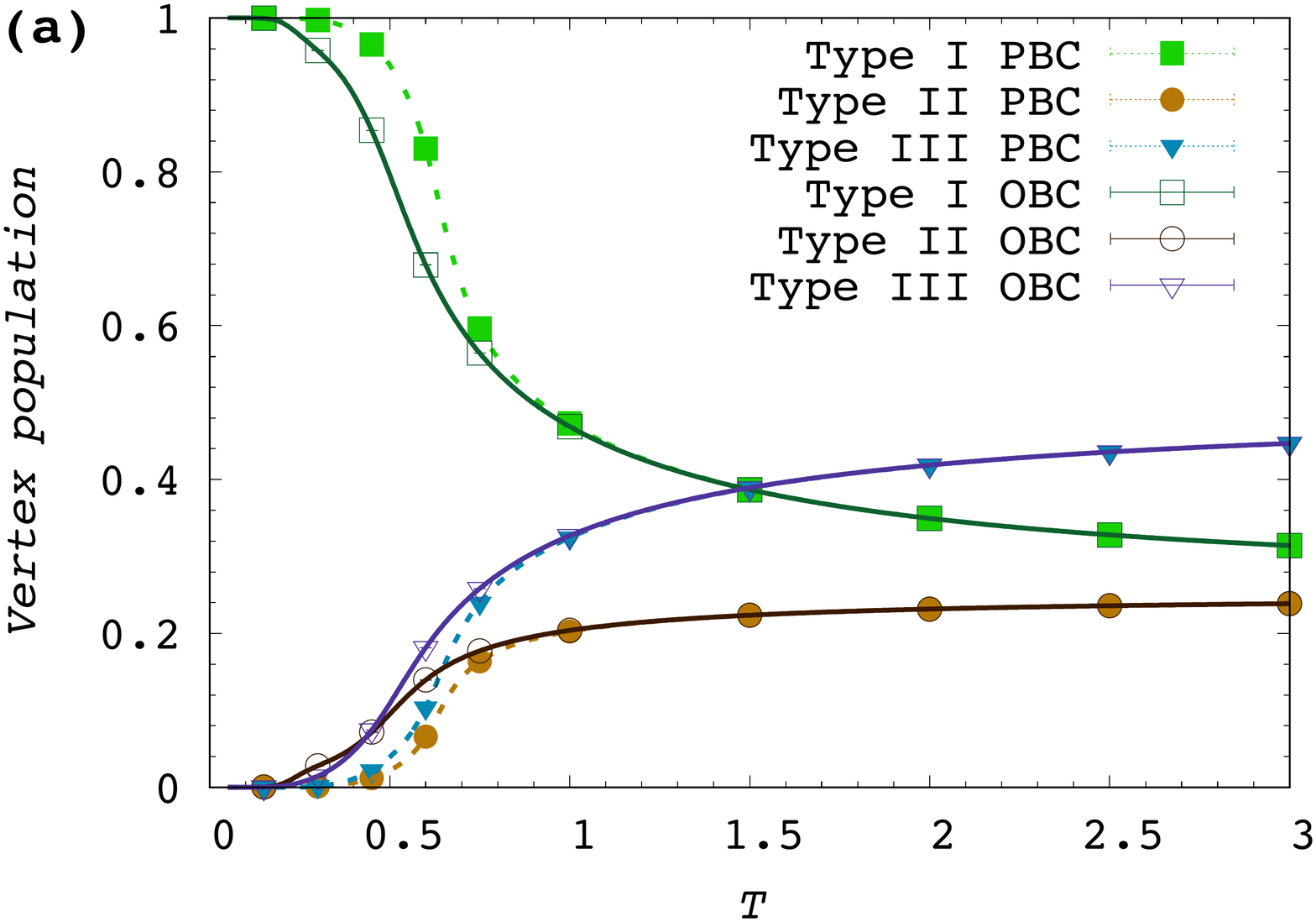} \hspace{.15\linewidth}
	\includegraphics[width=.4\linewidth]{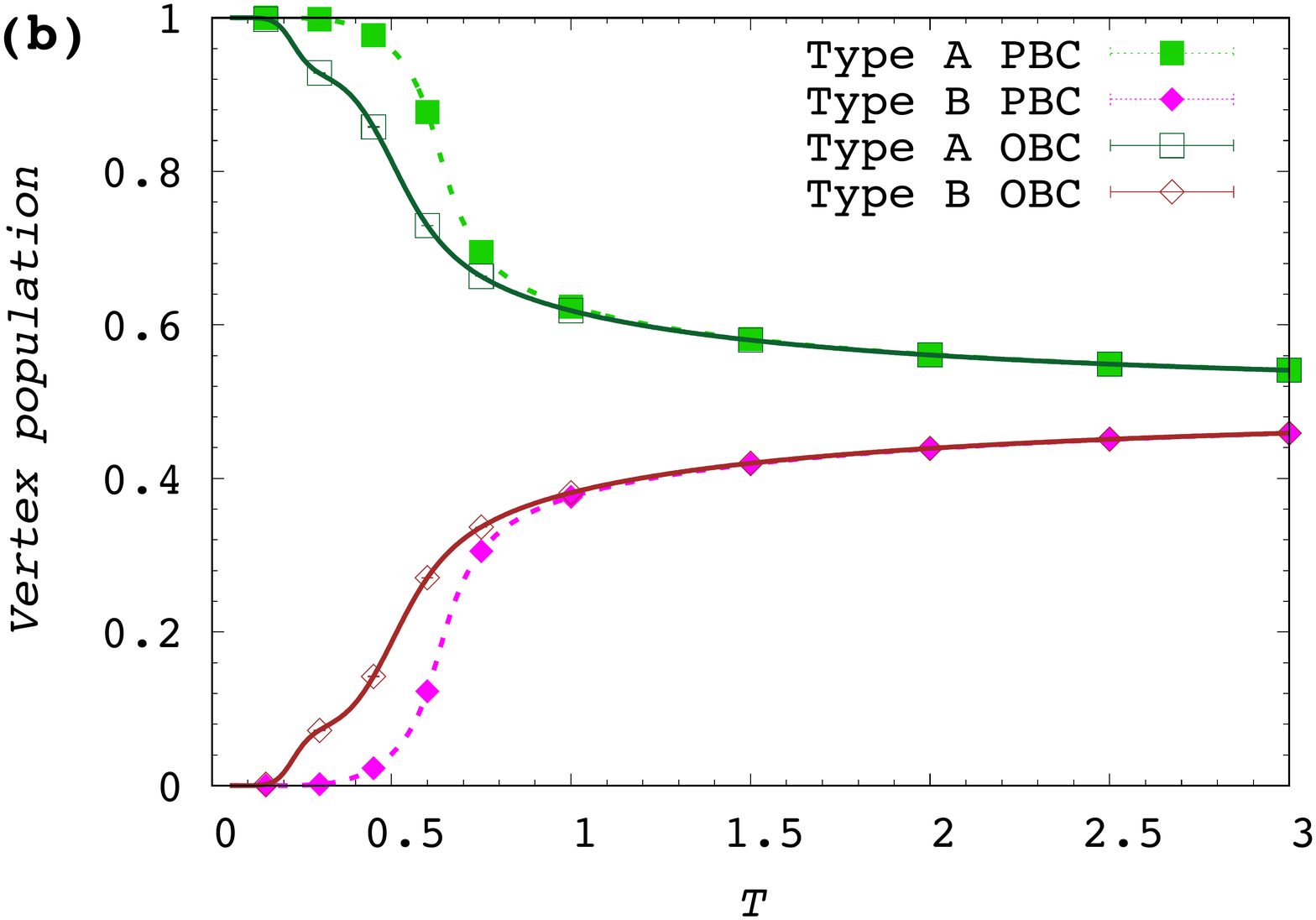}
    \includegraphics[width=.4\linewidth]{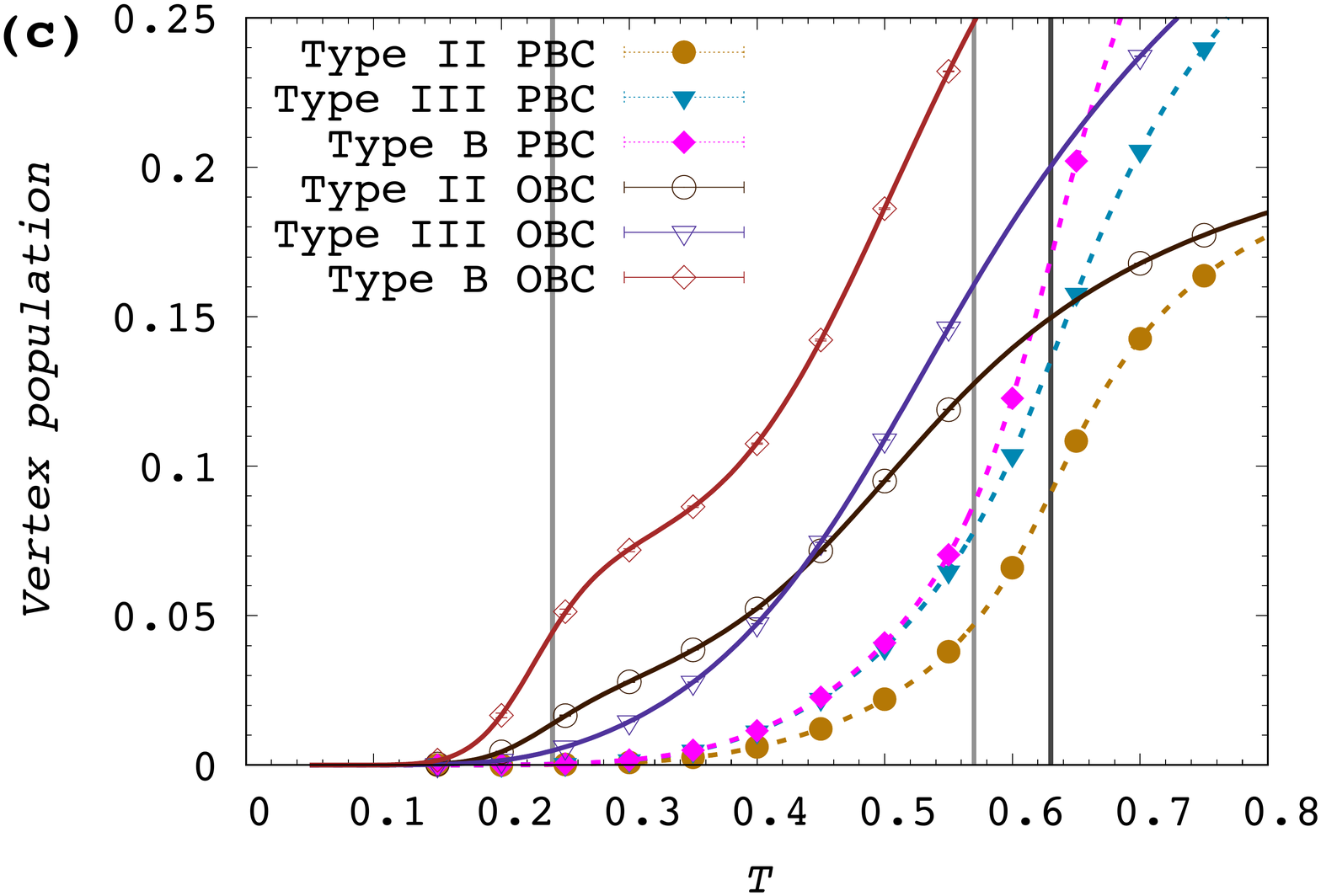}\hspace{.15\linewidth}
    \includegraphics[width=.4\linewidth]{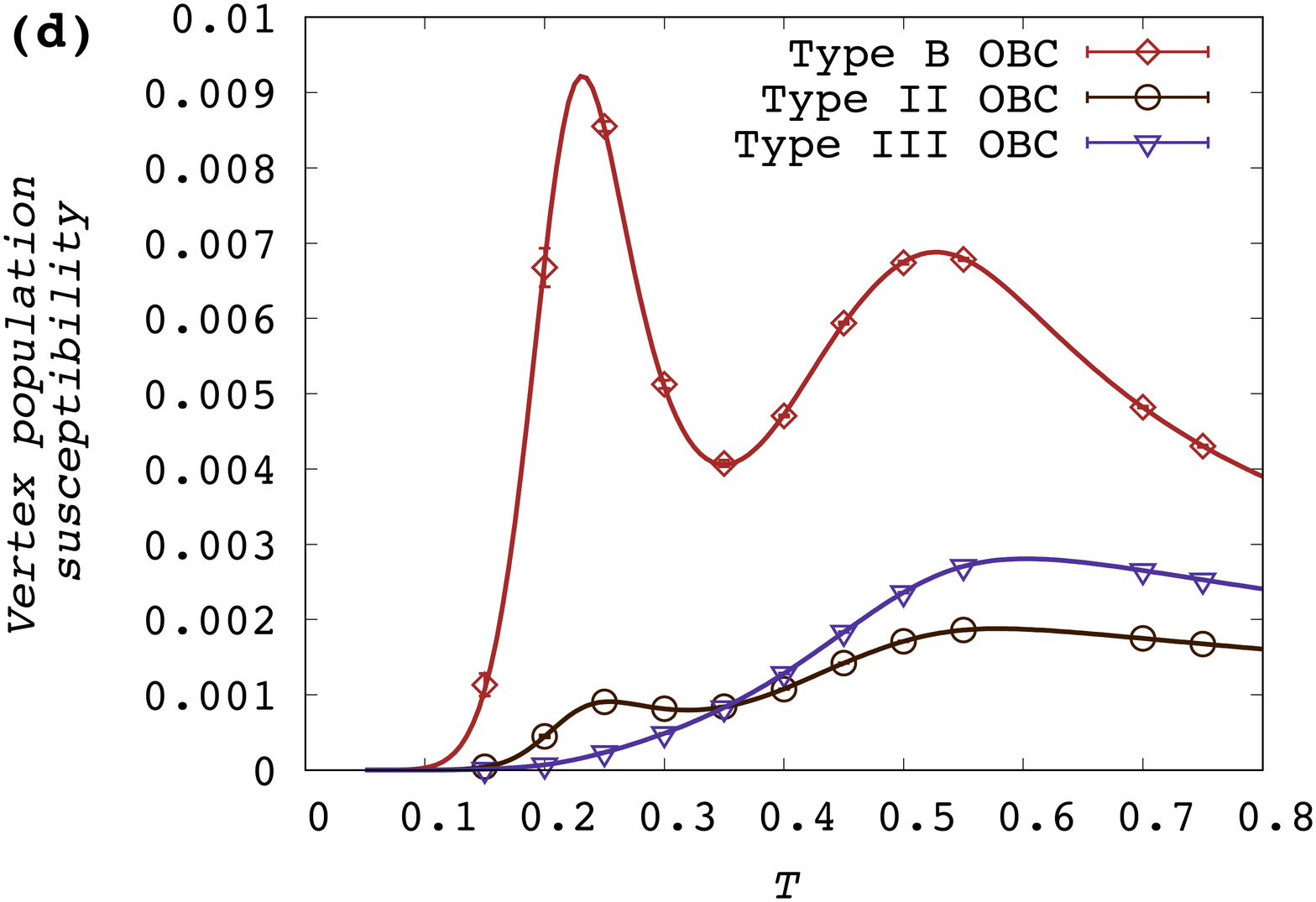}
    \caption{\textbf{(a)} Normalized vertex populations for the PBC and OBC lattices, $L=8$, as a function of temperature. \textbf{(b)} Secondary vertex populations as a function of temperature. Type B vertices appear at a lower temperature with OBC than with PBC. As the temperature rises, all vertex populations converge to their correspondent multiplicities. \textbf{(c)} Comparison of types II, III, and B populations for OBC and PBC. Vertical lines correspond to transition temperatures for $L=8$ with OBC (light gray) and PBC (dark gray). In the finite system, type II and type B vertices are slightly favored in the intermediate phase.  \textbf{(d)} The susceptibility of vertex populations (given in arbitrary units) with OBC shows two peaks for types II and B, and only one peak for type III. This extra peak appears as a consequence of the formation of domain barriers in the low-temperature transition.}
    \label{fig:vertex_compare}
\end{figure*}

To gain more insight on the mechanism that drives these different behaviors, we looked at the population of vertices that belong to each type as a function of temperature (see Fig. \ref{fig:vertex_compare}). The PBC system presents a trivial behavior in which low energy vertices (types I and A) are preferred at low temperature and higher energy vertices are increasingly common as the temperature rises. With OBC, however, we see that vertices of types II and B increase more rapidly in the region of the intermediate phase, as shown in Fig. \ref{fig:vertex_compare}(c). This is actually a requirement for the formation of the domain walls that characterize this phase, since these vertices are the building blocks of the wall.
We also computed the susceptibility of type II, type B, and type III vertex populations in the OBC lattice (Fig. \ref{fig:vertex_compare}(d)) to show that the first two exhibit strong peaks at both transition temperatures while the latter presents only one peak at the high-temperature transition, denoting that types II and B are indeed the most important excitations that take part in the low-temperature transition. Type I and type A vertices also present a peak in susceptibility (not shown here) as a consequence of more vertices assuming types II and B configurations, respectively. For PBC, only one peak in susceptibility is observed for all vertex types (also not shown), which coincides with the transition temperature as expected.

It is important to remark
that in order to transition from a typical intermediate state configuration to the ground state by means of single spin flips, the system has to access states of significantly higher energy in its configuration space. This important energy barrier poses a difficulty to obtaining the ground state through an annealing process. In fact, in our simulations the ground state has only been reached when we added multiple spin flips that attempt to invert either half of the lattice or entire lines of spins at once. As a consequence, a similar annealing process in an experimental setting with relatively small lattices would presumably result in many samples being found in a frozen-in intermediate phase even at very low temperatures. 

In order to understand the different behaviors observed for PBC and OBC in the present study, it is useful to resort to an analogy with ferromagnetism in continuous media. Firstly, it is important to notice that the ferromagnetic ground state with both PBC and OBC is induced by the head-to-tail arrangement of the neighboring dipoles, which is permitted by the lattice geometry. In the lattice with OBC, though, this fully magnetized state gives rise to a significant stray field, whose energy tends to be reduced by the formation of domains (see, for example, Ref. \onlinecite{Aharoni-book}). This domain formation, however, imposes an extra energy cost associated with domain boundaries, so that the total energy of a certain state is the result of a competition between stray field reduction and domain boundary formation energy. In the RWASI lattice, at least for the lattice sizes we were able to consider, this domain formation actually increases the total energy when compared to the fully magnetized state, in such a way that the ground state remains the same. Nevertheless, the energy difference between the fully magnetized state and the domain-wall state portrayed in Fig. \ref{fig:fase_inter} is certainly much smaller for OBC than for PBC, because the formation of domains with periodic conditions presents no advantage in terms of surface charge cancellation and stray field reduction. Moreover, the flipping of spins pertaining to the domain wall in the configuration shown in Fig. \ref{fig:fase_inter} imposes very little energy costs, since their interactions with neighbors are not all satisfied. This results in a great number of slightly different domain-wall states possessing similar internal energy, contrary to what occurs in the ground state, in which single-spin flips are generally costly. For a certain interval of temperatures, this combination of energetic and entropic effects reduces the free energy of the system favoring the domain-wall states and engendering the appearance of the intermediate phase observed in the simulations.

Another way to see that the domain wall formation is energetically disfavored with PBC is by noticing that the periodic conditions introduce topological constraints that are absent in the free boundary lattice. For instance, in the configuration shown in Fig. \ref{fig:fase_inter}, when PBC are considered a second domain wall is identified in the upper/lower boundary, doubling (at least) the energy associated with the domain wall and preventing the formation of domains. Furthermore, spin flips on the lattice edges with OBC costs significantly less energy than spin flips in the bulk, suggesting that the emergence of excitations required for domain wall formation is much more likely to occur for OBC than for PBC.

The effects of open boundaries on nanomagnetic arrays have also been discussed by Mac\^edo \textit{et al.} \cite{Macedo2018} for the pinwheel artificial spin ice, where it was shown that for OBC the ground state is composed by domains that are not present in the system with PBC. In addition, they found that different domain structures appear depending on the lattice shape and edge types. In this way, one may expect to find different domain structures or even the enhancement or suppression of the intermediate phase we have found by considering different shapes and edge types for the OBC lattice. An important difference between the two systems, however, is that in the pinwheel ASI the formation of domains actually changes the ground state for OBC, while in the RWASI the ground state for PBC and OBC coincide for the lattice sizes considered in this work. We hypothesize that this difference may be due to the greater stray field reduction provided by the flux closure domains observed in the pinwheel ASI, which are geometrically prohibited in the RWASI.

\section{Conclusion}

In summary, we have investigated the thermodyamic properties of a artificial spin ice geometry that had been proposed in Ref. \onlinecite{Wang2016} by means of Monte Carlo simulations. The prospects of technological applications related to the possibility of total control over the system's microstates allowed by this type of ASI, as well as its suitability for experimental studies addressing the equilibrium and non-equilibrium properties of magnetic arrays of nanoislands, justify the exploration of its thermal properties.

By analyzing low energy excitations in the system, we see that the magnetic monopole picture of the square ASI\cite{Mol2009a} is present in this geometry, even though the local frustration is absent. In addition, we observe that the secondary vertices of the RWASI lattice also support the presence of monopole-like excitations, which are connected by an energetic string and whose interactions are characterized by a Coulombic term. These secondary monopole-like excitations (type B monopoles) were found to have a slightly smaller magnetic charge, $q_B=1.21(1)Dl$, when compared to the monopoles in the main vertices (type III monopoles), $q_{III}=1.65(2)Dl$. This difference might be associated with the misalignment of the spins that form type B vertices.
Although the RWASI geometry constrains monopoles to move along straight lines, no signals of dimensional reduction were found.

By using a method based on the zeros of the energy probability distribution (EPD)\cite{Costa2017,Mol2018,Costa2019}, the transition temperature for periodic boundary conditions was obtained ($T_c^*=0.671(1)D/k_B$). In this case, a single phase transition between the ordered ground state and a paramagnetic state was found, and only above the transition temperature the density of monopoles increase considerably. This phase transition was characterized as belonging to the 2D Ising universality class, as one would expect by using universality arguments and comparing with the square ASI\cite{Silva2012}.

Our results for open boundary conditions, however, show a different and much more interesting behavior. An intermediate phase characterized by low magnetization and the presence of large domains of magnetized regions was observed. As a consequence, a leading zero of the EPD indicating a low temperature pseudo phase transition appears. This pseudo transition seems to be first order in nature as indicated by the Binder's cumulant of the magnetization. Moreover, the domain walls are composed mainly by type B and type II vertices, suggesting that a current of magnetic monopoles may be expected to develop in the formation of such domain walls. This expectation is confirmed by a peak in the susceptibility of the populations of type B and type II vertices. We remark that monopoles are constrained to move along the $\hatb{x}$ direction, in a way that those possible magnetic currents may be more easily measured.

Another point that deserves attention is that this low temperature pseudo transition is expected to be suppressed in the thermodynamic limit. However, as the simulated system sizes are on the same order of magnitude as the size of some typical realizations of ASI, we expect that the transition will play a determinant role in experimental realizations of this system. Indeed, considering $l=300nm$ the lateral dimension of the largest system that we simulate ($L=28$) would be about $30\mu m$, whereas the samples of Ref.\onlinecite{Wang2016} are approximately $80\mu m$ wide. Since the difference in behavior observed for OBC and PBC is a result of the energetic and entropic contributions of the domain formation process induced by the array edges, we expect the shape and size of the sample to give rise to different domain structures and significantly influence the intermediate phase.

\appendix*

\section{EPD zeros}
Here we present the main ingredients necessary to understand the EPD zeros scheme. For further details we refer the reader to Refs. \onlinecite{Costa2017,Mol2018,Costa2019}. Let us start by writing the partition function as
\begin{equation}
	\label{eq:partition7}
	Z(\beta)=\sum_E g(E)e^{-\beta E}=\sum_E h_0(E)e^{-\Delta \beta E},
\end{equation}
where the summation runs over all possible energy values $E$, $g(E)$ is the density of states, $\Delta \beta=\beta-\beta_0$, and $h_0(E)=g(E)e^{-\beta_0 E}$ is the non-normalized energy probability distribution---or, in other words, the histogram at an inverse temperature $\beta_0$. By assuming a discrete set of energies $E_n=\epsilon_0+n\epsilon$ and defining $x\equiv e^{-\Delta \beta \epsilon}$, the partition function can be rewritten in polynomial form:
\begin{equation}
	\label{eq:partition9}
	Z=e^{-\Delta \beta \epsilon_0} \sum_n h_0(n)x^{n},
\end{equation}
The zeros of this polynomial are  complex numbers. Since all polynomial coefficients, $e^{-\Delta \beta \epsilon_0}h_0(n)$, are real positive numbers, there are no real positive zeros and all roots appears as conjugated pairs. However, in the thermodynamic limit, one expects to find a zero at the point $(1,0)$ in the complex plane for $\beta_0=\beta_c$, signaling the non-analytic behavior of the free energy ($f=-k_BT\ln(Z)$) at the phase transition temperature, $\beta_c$. For finite systems,  the zero that is closest to this point is considered the dominant zero, and its real part can be used to estimate the critical temperature through the expression:
\begin{equation}
	\label{eq:iterazero}
	\beta_c=\beta_0-\frac{\ln ( \Re e (x_c) )}{\epsilon},
\end{equation}
where $x_c$ is the dominant zero at inverse temperature $\beta_0$. By making a new histogram at $\beta_0=\beta_c$, we can further refine our estimate for the critical temperature, making this an iterative process. It is also worth noticing that the imaginary part of the dominant zero is expected to scale with system size $L$ as
\begin{equation}
	\label{eq:imag}
	\Im m(x_c) \sim L^{-1/\nu},
\end{equation}
where $\nu$ is the critical exponent defined by the divergence of the correlation length.

\begin{acknowledgements}
We would like to thank B.V. Costa and J.C.S. Rocha for valuable discussions. The authors thank CAPES, FAPEMIG and CNPq for financial support.
\end{acknowledgements}

\bibliography{references}

\end{document}